\documentclass[10pt]{article}

\usepackage{amsmath}
\usepackage{amssymb}

\usepackage[sc]{mathpazo}
\usepackage[T1]{fontenc}

\usepackage{graphicx}

\usepackage{cite}

\usepackage{color} 

\usepackage[labelfont=bf,labelsep=period,justification=raggedright]{caption}
\usepackage{url}

   \graphicspath{{./figures/}{./ieee_sensors/figures/}}
   \DeclareGraphicsExtensions{.pdf,.jpeg,.png,.jpg}
\usepackage{amsmath,amssymb}
\newcommand{\bm}[1]{\boldsymbol{#1}}    

\def\Npdf{{\mathcal N}}		

\def\E{\bm {\mathcal{E}}}

\makeatletter
\renewcommand{\@biblabel}[1]{\quad#1.}
\makeatother

\usepackage{caption}
\usepackage{subcaption}
\usepackage[font=footnotesize,caption=false]{subfig}
\usepackage{cite}

\date{}

\begin{document}

\begin{center}
{\Large
\textbf{The Smartphone Brain Scanner: \\ A Mobile Real-time Neuroimaging System}
}
\\[0.2in]
Arkadiusz~Stopczynski, 
Carsten~Stahlhut, 
Jakob~Eg~Larsen,
Michael~Kai~Petersen,
Lars~Kai~Hansen
\\[0.05in]
arks@dtu.dk, csta@dtu.dk, jaeg@dtu.dk, mkai@dtu.dk, lkai@dtu.dk
\\[0.05in]
Section for Cognitive Systems, DTU Compute, Technical University of Denmark, Building 303B, DK-2800 Kgs. Lyngby, Denmark
\end{center}

\section{Introduction}

In the last few years, the research communities that study human behavior have gained access to unprecedented computational and sensing power that basically "fits into a pocket". This has happened for both specialized equipment used for building research tools, such as Reality Mining Badges \cite{choudhury2003sensing} or accelerometer sensors \cite{van2000shall}, as well as for consumer-grade, off-the-shelf devices. Smartphones and tablets are capable of sensing, processing, transmitting, and presenting information. This has already had a significant impact on many research domains, for example social science \cite{aharony2011social}, computer human interaction \cite{brown2011into}, or mobile sensing \cite{jensen2010estimating,kwok2009personal}. In neuroscience there is a widely recognized need for mobility, i.e.,~for  devices that support quantitative measurements in natural settings  \cite{makeig2009linking,blankertz2010berlin,gramann2011cognition}. Here we present our work on the \emph{Smartphone Brain Scanner} investigating the feasibility of off-the-shelf, consumer grade equipment in a neuroscience context, building a mobile real-time platform for stimulus delivery, data acquisition, and processing with a focus on \emph{real-time imaging} of brain activity.

Consumer grade neuroheadsets, capable of recording brain activity generated by post-synaptic potentials of firing neurons, captured through electrodes placed on the scalp using Electroencepahlography (EEG), have only recently made mobile brain monitoring feasible. Seen from a mental state decoding perspective, even a single channel EEG recording measuring the changes in electrical potentials, based on a passive dry electrode positioned at the forehead and a reference typically placed on the  earlobe, allows for measuring mental concentration and drowsiness, by assessing the relative distribution of frequencies in brain wave patterns throughout the day. Or, simply measuring the dynamic variability of brain wave frequency components in a mobile scenario, may be translated into neural signatures, e.g. reflecting whether a user is on the phone while driving a car \cite{yasui2009brainwave}. Similarly, positioning the single EEG electrode headband over the temple, may provide the foundation for building a Brain Machine Interface (BCI) utilizing the ability to capture steady state visual evoked potentials (SSVEP) from the visual cortex when looking at flashing lights patterns, and thereby design a BCI interface for prediction with high accuracy and no previous training when a disabled user is focusing on a specific area of a screen, based on the time locked EEG traces automatically generated as multiples of the particular flashing light frequencies \cite{luo2010user}.

As an example of the underlying technology used in several consumer products, the ThinkGear module manufactured by NeuroSky \footnote{http://www.neurosky.com/Products/ThinkGearAM.aspx} integrates a single dry electrode, reference and ground, attached to a headband. Essentially a system on a chip, it providies A/D conversion and amplification of one EEG channel, capable of capturing brain wave patterns in the 3-100 Hz frequency range, recorded at 512Hz sampling rate. Consumer neuroheadsets such as those manufactured by Emotiv \footnote{http://www.emotiv.com} provide low density neuroimaging based on 16 electrodes and typically support real-time signal processing in order to complement standard EEG measures with aggregate signals, that provide additional information on changes in mental state, or facilitate control of peripheral devices related to games. Their portability and built-in wireless transmission make them suitable for development of fully mobile systems that allow for running EEG experiments in natural settings. The improved comfort of these mobile solutions also allows for extending neuroimaging experiments over several hours. Furthermore, the relatively low cost of the neuroheadsets and mobile devices potentially open new opportunities for conducting novel types of social neuroscience experiments, where multiple subjects are monitored while they interact \cite{konvalinka2012two,dumas2011towards}.

However such 'low-fi' mobile systems present a number of challenges. In real-time applications, requiring signal processing to be performed with the lowest possible delay in order to present feedback to the user, the limited computational power of mobile devices may be a constraint. A solution might  be to offload parts of the processing to an external server and retrieve the processed results over the network. Consumer-grade mobile devices also present technical challenges for writing high-quality software: the devices operate on non-real-time operating systems ill-suited for time-sensitive tasks. These limitations might also affect timing of visual or auditory stimuli presentation, as well as synchronization with other sensors. From a neuroscience perspective, both the low-resolution recordings and artifacts induced in a mobile setup present significant challenges. Noise and confounds are introduced by movement of the subject and electrical discharges, while the positioning of the electrodes might be less than ideal when compared to a standard EEG setup\cite{stahlhutevaluation,Chi2012:EEGSensorsMobileBCI}. Nevertheless, we hold that these drawbacks  are clearly offset by the advantages of being able to conduct studies incorporating larger groups of subjects over extended periods of time in more natural settings. We suggest that  mobile EEG systems can be considered from two viewpoints: As stand-alone portable low-fi neuroimaging solutions, or alternatively as an add-on for retrieving neuroimaging data under natural conditions complementary to standard neuroimaging lab environments. 

In terms of software programming, creating a framework for applications in C++ rather than prevalent environments such as MATLAB, while approaching the problem as a smartphone sensing challenge, might enable new types of contributions to neuroscience. The Human Computer Interaction (HCI) community is already starting to apply consumer-grade headsets  to extend existing paradigms \cite{vi2012detecting}, thus incorporating neuroscience as a means to enhance data processing. Similarly the availability of low cost equipment means that even  general 'hacker and tinkerers' audiences gain interest in using neuroscience tools \footnote{http://neurogadget.com}. We see a great value in the emerging potential of entirely new groups of researchers and developers getting interested in neuroscience and obtaining tools allowing them to develop new kinds of applications. 

\section{Related Work}
Our real-time imaging EEG setup mediates between to hitherto disparate fields in sensorics, being on one hand a down-sized neuroimaging device and on the other hand a sophisticated smartphone sensor system for cognitive monitoring in natural conditions. We therefore briefly review the state of the art in both domains.

\subsection{Neuroimaging}

Several software packages for off-line and on-line analysis of biomedical and EEG signals are available. The most popular packages for off-line analysis are EEGLAB and FieldTrip; for building real-time BCI-oriented applications, notable frameworks are BCILAB, OpenViBE, and BCI2000.

EEGLAB is a toolbox for the MATLAB environment for processing collections of single-trial or averaged EEG data\cite{delorme2004eeglab}. Functions available in this framework include data importing, preprocessing (artifact rejection, filtering), independent component analysis (ICA), and other.
The framework can be used via a graphical interface or by directly manipulating MATLAB functions. The toolbox is available as open source (GNU license) and can be extended to incorporate various EEG data formats coming from different hardware.
Similarly, FieldTrip is an open source (GNU License) MATLAB toolbox for the analysis of MEG, EEG, and other electrophysiological data \cite{oostenveld2011fieldtrip}. Among others, FieldTrip has pioneered high-quality source reconstruction methods for EEG imaging. Fieldtrip has support for real-time processing of data based on a buffer construction that allows chunking of data for further processing in the MATLAB environment. 

BCILAB is a toolbox for building online brain-computer interface (BCI) models from available data \cite{delorme2011eeglab}. It is a plugin for EEGLAB running in MATLAB, providing functionalities for designing, learning, use, and evaluation of real-time predictive models. BCILAB is focused on operating in real-time  for detecting and classifying cognitive state. The classifier output from BCILAB can be streamed to a real-time application to effect stimulus or prosthetic control, or may be derived post-hoc from recorded data. The framework is extensible in various layers: additional EEG hardware as well as data processing steps (e.g.~filters and classifiers) can be added. But as these tool-boxes are developed within the MATLAB environment, neither FieldTrip's real-time buffer nor BCILAB are suitable for mobile application development.

OpenViBE is a software framework for designing, testing, and using Brain Computer Interfaces \cite{renard2010openvibe}. The main application fields of OpenViBE are medical i.e. assistive technologies, bio- and neurofeedback as well as virtual reality multimedia applications . OpenViBE is open source (LGPL 2.1) and targets an audience focused on building real-time applications for  Windows and Linux Operating Systems, and does not specifically support light-weight mobile platforms.  A similar C++ based framework for building real-time BCI applications is BCI2000 \cite{schalk2004bci2000}. A comprehensive review of the BCI frameworks can be found in \cite{brunner2011bci}. Some of the consumer EEG systems also include Software Development Kits (SDKs) allowing for data acquisition, processing, and building applications.
Emotiv SDK, available with Research Edition of Emotiv system is multi-platform, currently running on Windows and OSX, with Linux support in beta. The SDK allows for building applications either using raw EEG data or extracted features including affective state and recognition of facial expressions based on eye movements. The extracted features can be integrated into a C++/C\# application through a set of dynamically linked libraries. Although such SDK frameworks can greatly speed up the process of building BCI applications, they are mostly targeted towards scenarios where immediate feedback is available such as gaming, and it remains a challenge to validate or tweak code for custom needs. To sum up, none of the aforementioned software platforms can easily be adapted to support mobile and embedded devices.

	\subsection{Cognitive Monitoring Systems}
Mobile brain imaging might also be viewed as yet another sensor extension to self-tracking applications, which have become prevalent with smartphones and the emergence of low-cost wearable devices lowering the barriers for people to engage in life logging activities\cite{swan2012sensor}. With the availability of multiple embedded sensors modern smartphones have become a platform for out-of-the-box data acquisition of mobility (GPS, cellular network, WiFI), activity level (accelerometer), social interaction (Bluetooth, call, and text logs), and environmental context (microphone, camera, light sensor)\cite{aharony2011social}.Recently non-invasive recording of brain activity has become common as several low-cost commercial EEG neuro-headset and headband systems have been made available, including apart from the previously mentioned Emotiv EPOC and NeuroSky \footnote{http://www.neurosky.com/}, the InteraXon Muse \footnote{http://www.interaxon.ca/}, Axio  \footnote{http://www.axioinc.com/}, and Zeo \footnote{http://myzeo.com/}. These sensors support applications ranging from BCI, game control, stress reduction, cognitive training, to sleep monitoring. These neuroheadsets feature up to 16 electrodes, but ongoing developments promise next-generation low-cost EEG devices with significantly higher number of electrodes, better quality signals, and improved comfort. The Smartphone Brain Scanner framework described in this paper, can be used with mobile EEG devices with various numbers of electrodes to allow for capture of neuroimaging data over several hours. Battery tests on Samsung Galaxy Note with all wireless radios and screen turned off resulted in 11 hours of uninterrupted recording and storage of data from an Emotiv EPOC headset. However in reality current generation neuroheadsets are limited by their solution-based electrodes which dry out, and more comfortable designs\cite{debener2012taking,looney2012ear} may be required for continuous mobile neuroimaging throughout the day.

Beyond EEG multiple bio signals and physiological parameters can contribute to cognitive state monitoring, such as  respiratory rate\cite{Moraveji2012}, heart rate variability, galvanic skin response\cite{poh2010wearable}, blood pressure, oxygen saturation, body/skin temperature, ECG, EMG, and body movements\cite{Pantelopoulos2010}. A webcam or a camera embedded in a smartphone allow measurements of heart rate, variability, and respiratory rate by analyzing the color channels in the video signal\cite{poh2011advancements}. Continuous monitoring of heart rate is enabled by pulse watches \footnote{http://www.polar.com/} and recently the Basis Band wrist-worn sensor \footnote{http://www.mybasis.com} that allow 24/7 recording under a subset of conditions (non-workout situations), allowing user mobility and measurements in natural conditions. The Q Sensor from Affectiva \footnote{http://www.affectiva.com/} is an example of a system for monitoring galvanic skin response (GSR) and accelerometer and temperature data from a wrist-worn device. FitBit \footnote{http://www.fitbit.com/} is an example of a wearable pedometer, monitoring number of steps taken, distance traveled, calories burned, and floors climbed.

\section{Methods: The Smartphone Brain Scanner}

\begin{figure}[!t]
\centering
\includegraphics[width=0.6\columnwidth]{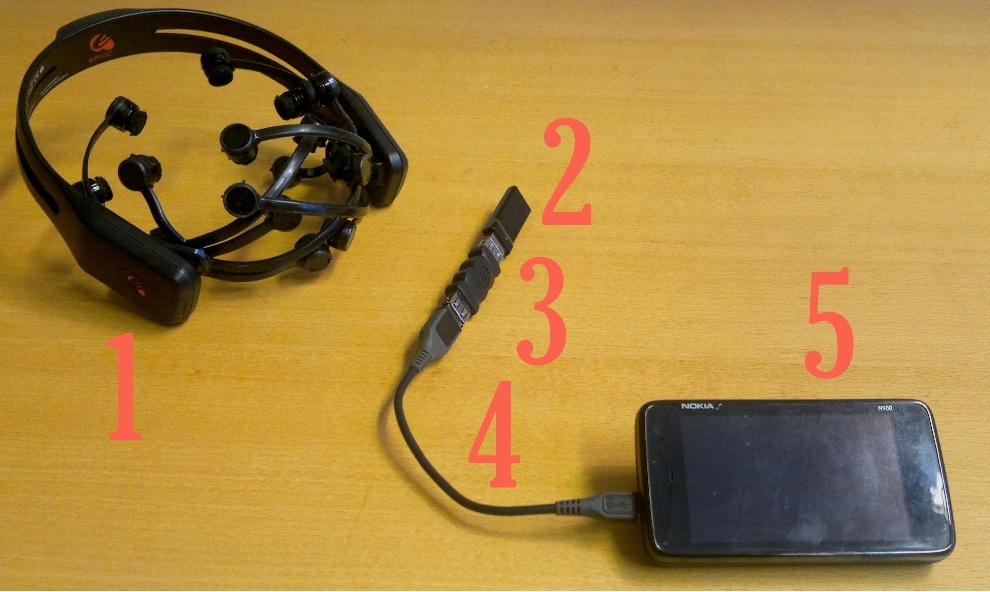}
\caption{The system with Emotiv EPOC wireless EEG headset (1), Receiver mod- ule with USB connector (2), USB connector and adapter (3+4), and Nokia N900 (5).}
\label{fig:sbs_system}
\end{figure}

The \emph{Smartphone Brain Scanner} (SBS2) is a software platform for building research and end-user oriented multi-platform EEG applications. The focus of the framework is on mobile devices (smartphones, tablets) and on consumer-grade (low-density and low-cost) mobile neurosystems (see Fig.~\ref{fig:sbs_system}. The SBS2 is freely available under the MIT License on GitHub at \url{https://github.com/SmartphoneBrainScanner}.

\begin{figure}[!t]
\centering
\includegraphics[width=1\columnwidth]{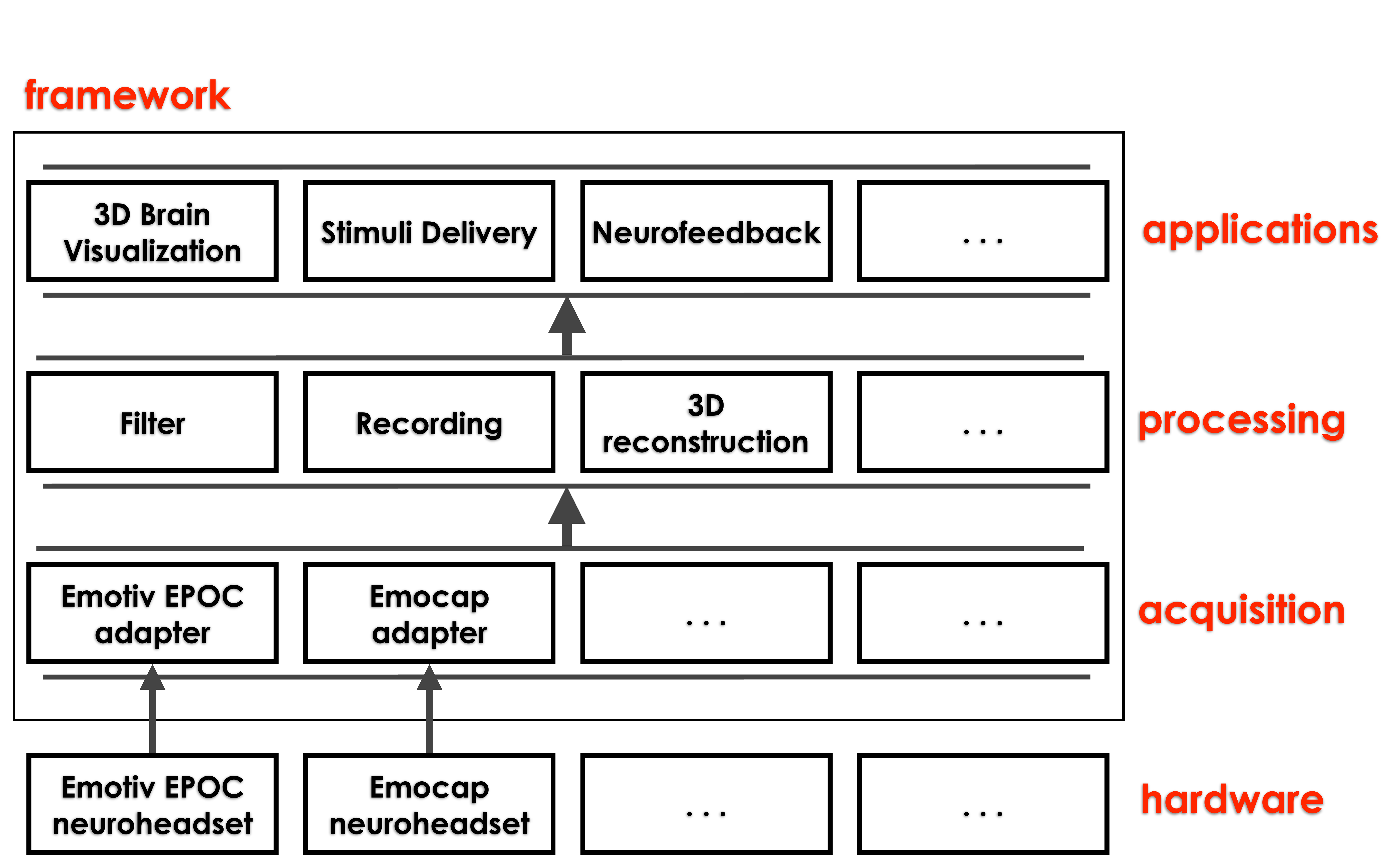}
\caption{Overview of the layered architecture of the SBS2 framework. Data from the connected EEG hardware are acquired and extracted by specific adapters and all subsequent processing is hardware agnostic. The empty boxes indicate the extensibility of the architecture allowing additional hardware devices for data acquisition and additional processing methods.}
\label{sbs2_arch_1}
\end{figure}

The SBS2 framework is divided into three layers: low-level data acquisition, data processing, and applications. The first two layers constitute the core of the system and include common elements used by various applications. An overview of the architecture is shown in Fig.~\ref{sbs2_arch_1}.

	\subsection{Key Features}
With focus on the mobile devices, SBS2 is a multi-platform framework. The underlying technology -- Qt -- is an extension of C++ and is currently supported on the main desktop operating systems (Linux, OSX, Windows) as well as mobile devices (Android, BB10, and partially iOS)\footnote{http://qt.digia.com/Product/Supported-Platforms/}.

We have aimed for a modular framework, allowing for adding and modifying data acquisition and processing blocks. The modules are created as C++ classes and integrate directly with the core of the framework. The framework supports building real-time applications; data can be recorded for subsequent off-line analysis, however most of the implemented data processing blocks aim to provide real-time functionality for working with the EEG signal. The applications developed with SBS2 can be installed on desktop and mobile devices, started by the user in a usual way and distributed via regular channels, such as repositories and application stores.

	\subsection{Data Acquisition}
\begin{figure}[!t]
\centering
\includegraphics[width=1\columnwidth]{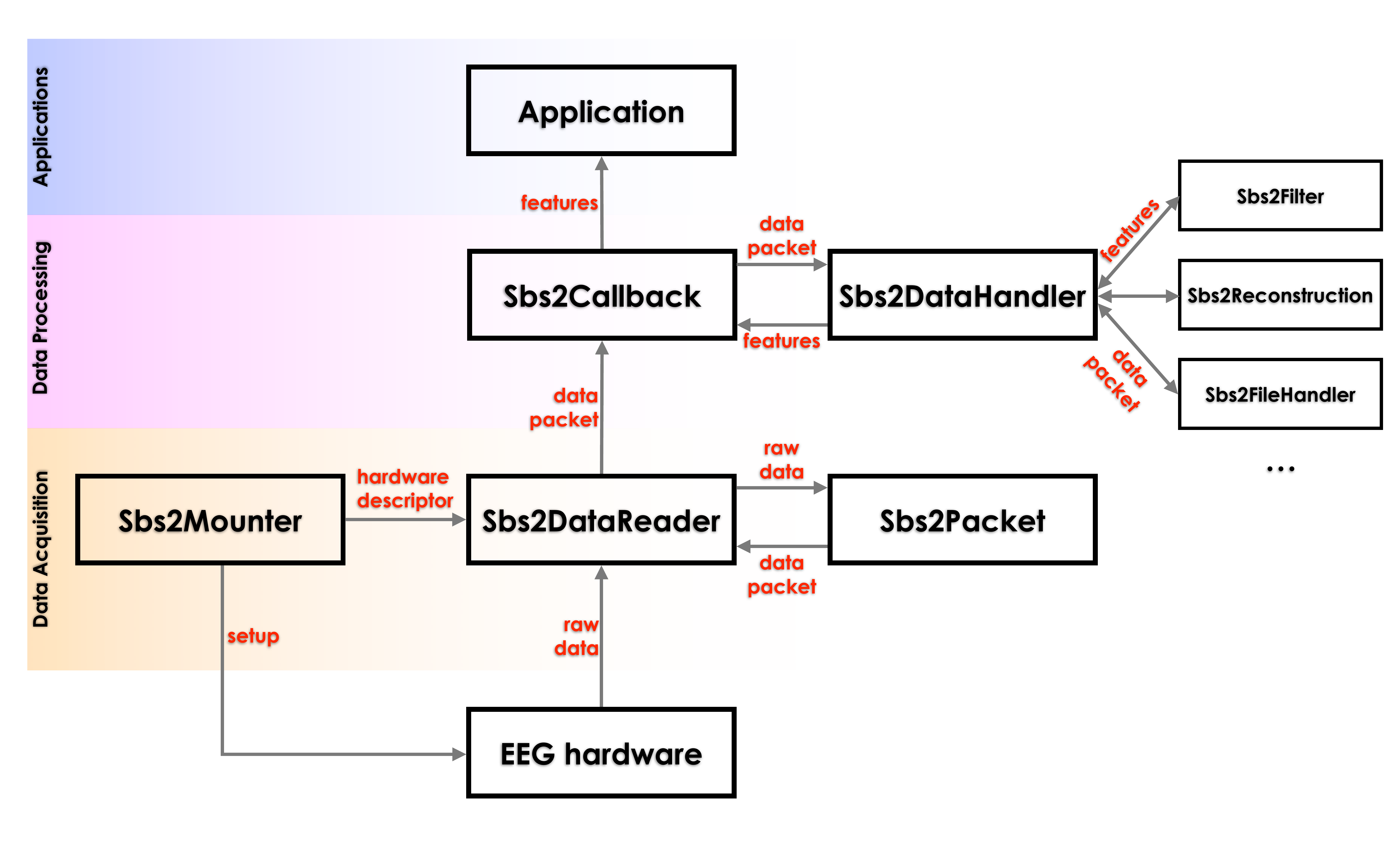}
\caption{The Smartphone Brain Scanner architecture. Data are acquired in the first layer from the EEG hardware, passed to the Data Processing Layer and extracted features as well as raw values are then available for applications.}
\label{sbs2_arch_2}
\end{figure}

The Data Acquisition layer is responsible for setting up communication with an EEG device, acquiring the raw data, and forming packets. Three primary objects are used: {\tt Sbs2Mounter}, {\tt Sbs2DataReader}, and {\tt Sbs2Packet}, thereby abstracting all the specificities of the EEG systems (hardware) and OS + device running the software (platform). Different embedded devices, even with the same OS may require specific code for certain low-level functionalities, for example to access the USB port.
A higher-fidelity architecture is shown in Figure \ref{sbs2_arch_2}. The EEG hardware is set up by a specialized {\tt Sbs2Mounter} object. The information about the hardware (e.g.~mounting point, serial number) is passed to a {\tt Sbs2DataReader} object. This object subsequently begins reading the raw data from the hardware. The raw data are passed to a {\tt Sbs2Packet} object to create a proper encapsulation, setting the values for all the EEG channels and metadata. Once formed, the packet is pushed to the Data Processing layer via a {\tt Sbs2Callback} object.

The Data Acquisition layer of the SBS2 is originally designed to support the Emotiv EPOC headset. It has been extended to support additional hardware, by implementing additional classes of the hardware mounter, data reader, and packet creator. For Emotiv headset, this layer also contains the data decryption module, as the stream coming from the device is encrypted.

Mounting the EEG hardware on a desktop and embedded devices require drivers, either standard kernel modules or proprietary drivers created by the vendor. The Emotiv EPOC USB receiver is mounted as \emph{/dev/hidraw} in Linux (desktop and Android), provided that the device and kernel support USB host mode and have the HIDRAW module enabled. Most desktop Linux flavors have both by default, but currently most Android mobile devices support only USB host mode out-of-the-box. In the current implementation a custom kernel needs to be compiled with the HIDRAW module enabled. Reading the data directly from the /dev/hidraw device requires 'root' privileges, which must be enabled on Android devices to acquire data from the Emotiv EPOC receiver. This is possible for most recent Android devices, e.g.~for the Nexus (developer) line of devices. We can expect that the next generation of mobile neuroheadsets will use standardized Bluetooth low-energy protocols and Android devices will be able to support them by default. This will likely have a significant impact on the adoption of neuroimaging outside lab environments.

	\subsection{Data Processing}
Well-formed EEG packet objects are used for data processing. The functionality of this layer is hardware agnostic and depends only on packet content, i.e. data for the EEG channels, reflecting a particular sensor configuration, and sampling frequency. Single packets are dispatched to different processing objects and methods, including recording, filtering, 3D reconstruction etc. Some operations need to collect data into frames and run asynchronously (in separate thread), pushing the results back to the callback object once the results are ready.

{\tt Sbs2Callback} is an object implementing the {\tt getData(Sbs2Packet*)} method, to which  single packets are always passed and can then be dispatched to the {\tt Sbs2DataHandler} or pushed to the Application layer. {\tt Sbs2DataHandler} is an object providing methods for data processing, by delegating them to specialized objects, including {\tt Sbs2FileHandler} and {\tt Sbs2Filter}.

The framework for data processing is extensible and new modules can be added to the core, the data handler prepares the data in a format expected by the processing block (e.g.~collecting packets into larger frames) and runs the processing method. The currently implemented blocks allow for a variety of processing operations. The raw EEG data can be recorded, including timestamped events (stimuli onsets, user responses etc.). Raw packets as well as extracted features and arbitrary values can be streamed over the network, either for data processing or for interconnection between devices, for example for multiplayer gaming. Other methods for data processing, including filter, FFT, spatial filter (CSP), and classifier (LDA) are also implemented and can be used for building the pipelines.

	\subsection{3D Imaging}

The most advanced data processing block of the Smartphone Brain Scanner is the source reconstruction aimed at real-time 3D imaging. Source reconstruction estimates the current sources within the brain that are most likely to have generated the observed EEG signal at the scalp level. As the number of possible source locations far exceeds the number of channels, this is known to be an extremely ill-posed inverse problem. A unique solution is obtained by imposing prior information in correspondence with e.g.~anatomical, physiological, or mathematical properties \cite{Baillet1997:BayesianAnatomFuncPrior,Phillips2002:AnaInformedBasis,Hamalainen1994:MNE}. Implemented inverse methods in the SBS2 covers Bayesian formulations of the widely used Minimum-norm method (MN) \cite{Hamalainen1994:MNE} and low resolution electromagnetic tomography (LORETA) \cite{PascualMarqui1994:LORETA}. The Bayesian formulation used in the SBS2 framework allows adaptation of hyper-parameters to different noise environments in real-time.
This is an improvement over previous real-time source reconstruction approaches \cite{Congedo2006:ClassificationByElectromagneticInverseSol,Noirhomme2008:SingleTrialEEGSourceReconBCI,Besserve2011:ImprovingQuantificationFunctionalNetworks} that apply heuristics to estimate the parameters involved in the inverse method.
The current source reconstruction is based on an assumed forward model matrix, $\bf A$, connecting scalp sensor signals ${\bf Y}$ (channel by time) and current sources ${\bf S}$ (cortical locations by time) \cite{Baillet2001:BrainMapping}
\begin{equation}\label{eq:ForwardProblem}
    {\bf Y} = {\bf AS} + {\E}.
\end{equation}
The term ${\E}$ accounts for noise not modeled by the linear generative model.
When estimating the forward model a number of issues are taken into consideration such as sensor positions, the geometry of the head model (spherical or 'realistic' geometry), and tissue conductivity values \cite{Wolters2007:NumercalMathematicsFEM,Hallez2007:reviewForwardProblem,Drechsler2009:FullSubtractionFEM}. With the  forward model $\bf A$ given and the linear relation in Eq.~\eqref{eq:ForwardProblem} the source generators can be estimated.
We assume the noise term to be normal distributed, uncorrelated, and time independent leading to the probabilistic formulation
\begin{eqnarray}\label{eq:Likelihood}
p \left( {{\bf Y} \left| {\bf S} \right. } \right) &=& { \prod_{t=1}^{N_t} {\Npdf} \left( {{\bf y}_t \left| {{\bf As}_t ,\beta^{-1}{\bf I}_{N_c } } \right.} \right)} \\
p \left( {\bf S} \right) &=& { \prod_{t=1}^{N_t} {\Npdf} \left( {{\bf s}_t \left| {{\bf 0} ,\alpha^{-1}{\bf L}^T {\bf L} } \right.} \right)}
\end{eqnarray}
Where $p \left( {\bf S} \right) $ is the prior distribution over $\bf S$ with $\bf L$ given as a graph Laplacian ensuring spatial coherence between sources, and $\beta^{-1}$ as the noise variance. Using Bayes' rule the posterior distribution over the sources is maximized by
\begin{eqnarray}\label{eq:MNsolution}
p \left( {{\bf S} \left| {\bf Y} \right. } \right) &=& { \prod_{t=1}^{N_t} {\Npdf} \left( {{\bf s}_t \left| {{\bm \mu}_t ,{\bf \Sigma }_{s} } \right.} \right)} \nonumber \\
{\bf \Sigma }_{s} &=& \alpha^{-1} {\bf I}_{N_d} - \alpha^{-1} {\bf A}^T {\bf \Sigma}_{y} {\bf A} \alpha^{-1}  \nonumber \\
{\bf \Sigma }_{y}^{-1}  &=&  {\alpha ^{ - 1} {\bf A}{\bf L}^{T}{\bf L}{\bf A}^{T}  + \beta ^{ - 1} {\bf I}_{N_c } }   \\
 {\bar{{\bf s}}_t}  &=& \alpha ^{ - 1} {\bf A}^{T} {\bf \Sigma }_{y} {\bf y}_t .
\end{eqnarray}
Here, $\bf L$ denotes a spatial coherence matrix, which in the current form take advantage of graph Laplacian using a fixed smoothness parameter ($0.2$). 
\section{Methods: Experimental Designs}
In this section we briefly describe the design of experiments that demonstrate and validate the potential of the SBS2 framework, the specific hardware, and the mobile approach in general.

	\subsection{Timing and Data Quality}

First, we analyze the data and timing quality. Many neuroscience paradigms rely heavily on accurate synchronization between EEG signal and stimuli, user response, or data from other sensors (e.g. P300, steady state visual evoked potentials). However, we can also envision applications in which the present 'low-cost' mobile setup will be used to collect data from many subjects over extended periods, where the precise synchronization is less important.

		\subsubsection{Emotiv EEG sampling}
The measurements are all based on the Emotiv EEG neuroheadset. The nominal sampling frequency of this neuroheadset is 128Hz (downsampled from internal 2048Hz). For validation purposes we test the actual sampling rate obtained from 3 randomly picked Emotiv devices (10 $\times$ 10 min measurements for each).
		\subsubsection{Data Quality}
The Emotiv hardware adds a modulo $129$ counter ($0-128$) to every packet transmitted from the device. This allows for data quality control (dropped packets) with accuracy of modulo $129$. It is possible to obtain long recordings (over one hour) using this neuroheadset and SBS2. The battery in the Emotiv hardware is rated at $12h$ of continuous operation; in recording-only setup mobile device such as Galaxy Note (offline mode, screen off, only decrypting and recording) lasts for around $10h$. Provided that good visibility between the Emotiv EEG neuroheadset transmitter (located in the back part of the headset) and USB receiver is maintained, we were able  to achieve zero packet loss in the full rundown recording.
In order to acquire an EEG signal of good quality, the impedance between the electrodes and the scalp should be kept under $5 k\Omega$. The Emotiv headset embeds the channel quality information in the signal directly (2Hz per channel, multiplexed into the signal). The values are unscaled, and come from applying a square wave of $128Hz$ to the DRL feedback circuit and extracting the amplitude of the inherent square wave using phase-locked detection on each channel. In principle the obtained values can be calibrated using a known impedance. For regular usage however, the hardware manufacturer assures that the green color of the indicator (channel quality value greater than 407) corresponds to sufficiently low impedance of the electrode. From our experience with the system this appears correct.
		\subsubsection{Timing}
\begin{figure}[!t]
\centering
\includegraphics[width=1\columnwidth]{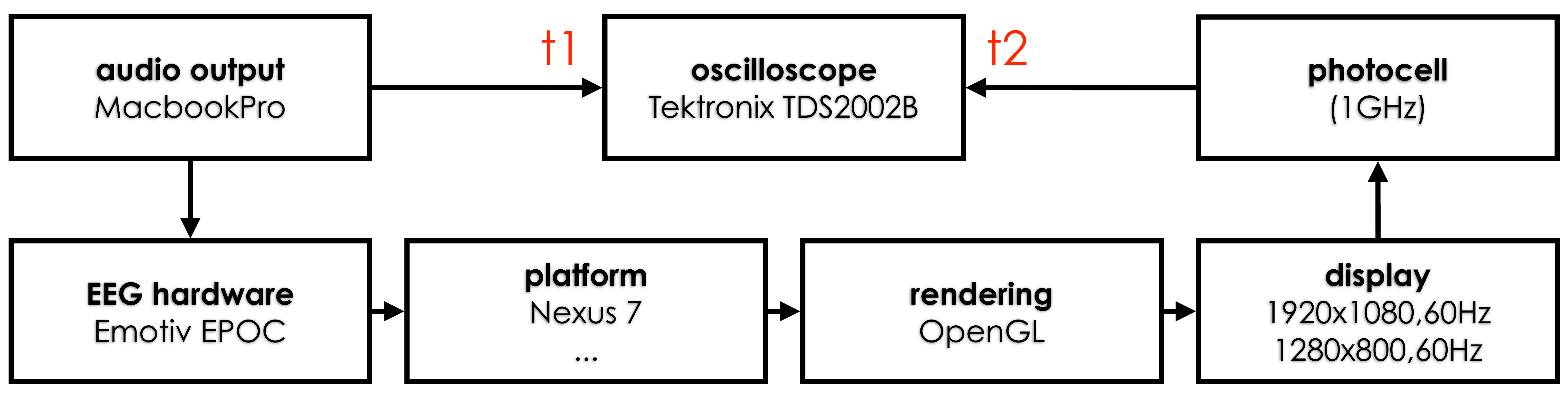}
\caption{The timing measurement setup. 10Hz sinusoid is generated with a sound card, amplified, and fed into an oscilloscope and the EEG hardware.}
\label{figure_timing_arch}
\end{figure}
In order to measure the total delay in the system, we use the setup as depicted in Fig.~\ref{figure_timing_arch}. A sinusoidal audio tone of $10Hz$ and trailing and following periods of silence is generated and amplified so it can be detected by the EEG hardware and split into oscilloscope and EEG hardware. The software on the device performs peak detection on the signal and visualizes the peaks by changing the screen color from black to white. This change is detected by a photocell, connected to the second channel of the oscilloscope. We can then calculate $dt1=t2-t1$, indicating the total delay of the system from the physical signal reaching the EEG hardware to it being visualized on the screen (without any additional processing), see Figure \ref{figure_timing_all}. We also look at the jitter $dt2$ as the difference between $min$ and $max$ values of $dt1$. The observed delta depends on the EEG sampling rate (here $128Hz$), the processing power of the device, and screen refresh rate ($60Hz$ for all tested devices).

	\subsection{Imagined Finger Tapping}
One of the best known and examined experiments from the BCI literature is a task in which a subject is instructed to select between two or more different imagined movements \cite{Muller-Gerking1999:DesigningOptimalSpatialFilters,Babiloni2000:LinearClassLowResEEGImagHand,Dornhege2004,Blankertz2006:BerlinBCI}. Such experiments are rooted in a central aim of many BCI systems, namely of being able to assist patients with severe motor disabilities to communicate by 'thought'.
In this contribution we replicate a classical experiment with imagined finger tapping (left vs. right) inspired by \cite{Blankertz2006:BerlinBCI}. The setup consisted of a set of three different images with instructions, \textit{Relax}, \textit{Left}, and \textit{Right}. In order to minimize the effect of eye movements, the subject was instructed to focus on the center of the screen, where the instructions also appeared (3.5 inch display size, 800 x 480 pixels resolution, at a distance of 0.5 m). The instructions \textit{Left} and \textit{Right} appeared in random order. A total of 200 trials were conducted for a single subject.

\section{Results and Discussion}
In this section we present and discuss the results of the experiments, validating the performance of the software, the used platforms, and EEG hardware.
	\subsection{Timing and Data Quality}
		\subsubsection{Emotiv EEG sampling}
\begin{figure}[!t]
\centering
\includegraphics[width=1\columnwidth]{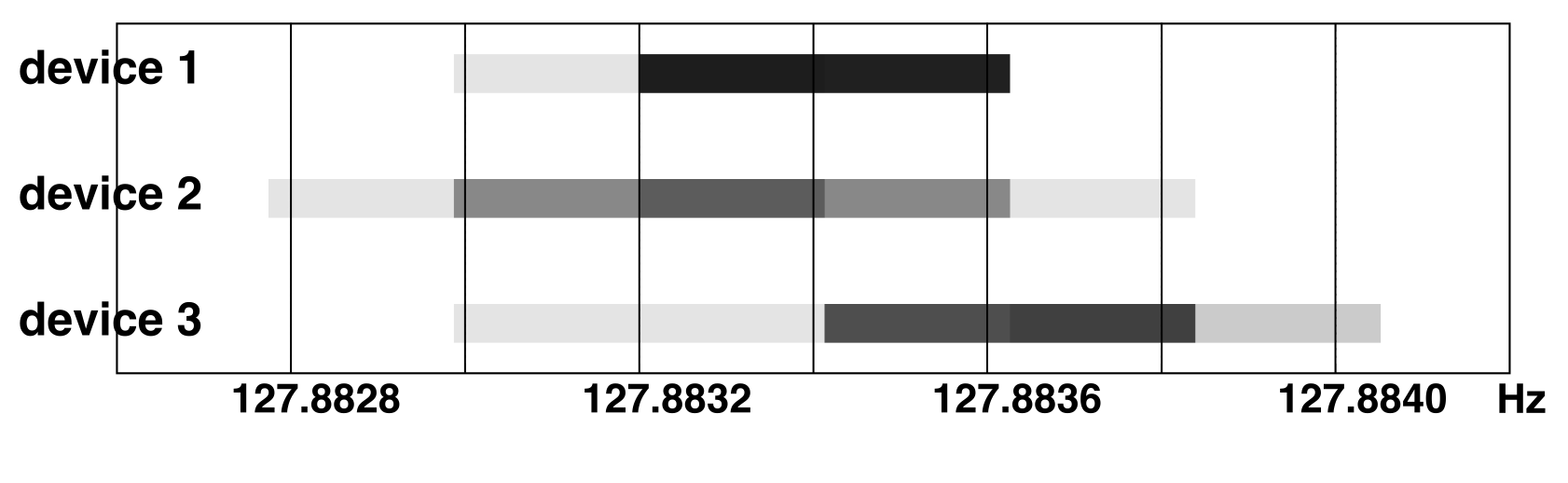}
\caption{Measured sampling frequency, including measurement resolution for 3 random Emotiv EPOC devices, $10\times 10min$ recordings for each.  All measured rates, including uncertainty are between $127.8828Hz$ and $127.8841Hz$, corresponding to $.99908$ and $.99909$ of nominal $128Hz$. The measurements were performed with $1ms$ resolution ($2ms$ accuracy) on $76800$ EEG packets. All tests were performed in a normal temperature on a single day.}
\label{figure_sampling}
\end{figure}

From Fig.~\ref{figure_sampling} we can see that the Emotiv EPOC hardware a) has an actual sampling rate close to $127.88Hz$ and b) keeps this sampling rate in a fairly consistent manner.  Depending on the analysis performed on the data, one can assume $128Hz$, $127.88Hz$, or measure the actual sampling rate for every Emotiv EPOC hardware device individually.
		\subsubsection{Timing}
\begin{figure}[!t]
\centering
\includegraphics[width=0.8\columnwidth]{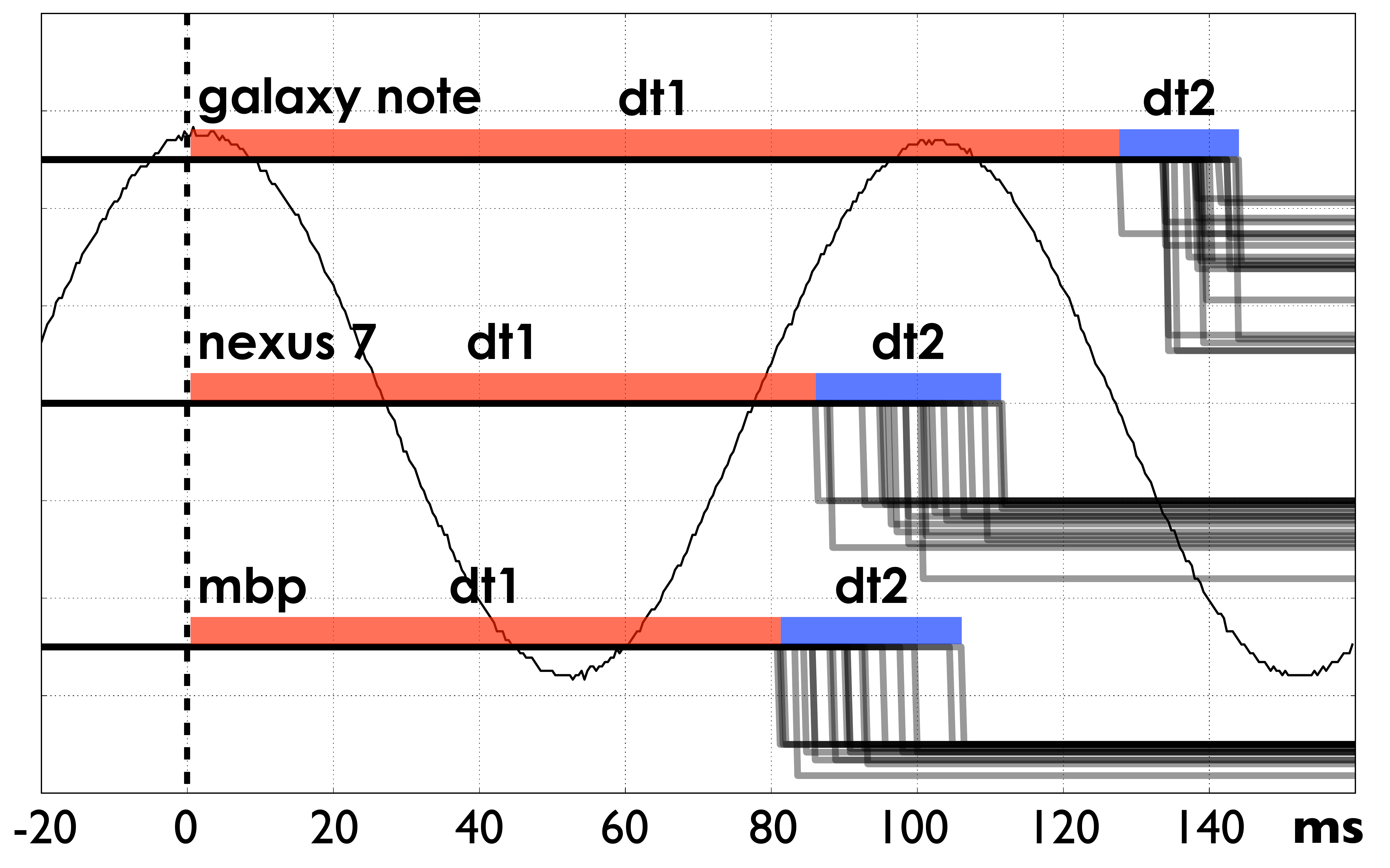}
\caption{System response timings. Galaxy Note running Android 4.0.1, 60Hz AMOLED screen, $t1=125ms$, $t2=16ms$; Nexus 7 running Android 4.1.1, 60Hz IPS LCD screen, $t1=85ms$, $t2=26ms$; MacbookPro, LCD screen (60Hz), $t1 = 80ms$, $t2=26ms$.}
\label{figure_timing_all}
\end{figure}

The results of the timing measurements (20 per device) are depicted in Fig.~\ref{figure_timing_all}.
We can see in the results that for all devices there is a significant delay between the signal reaching the EEG hardware and being fully processed in the software ($80-125ms$). This delay however, although significant is fairly stable ($16-26ms$ jitter) and thus can be corrected for.

\begin{figure}[!t]
\centering
\includegraphics[width=1\columnwidth]{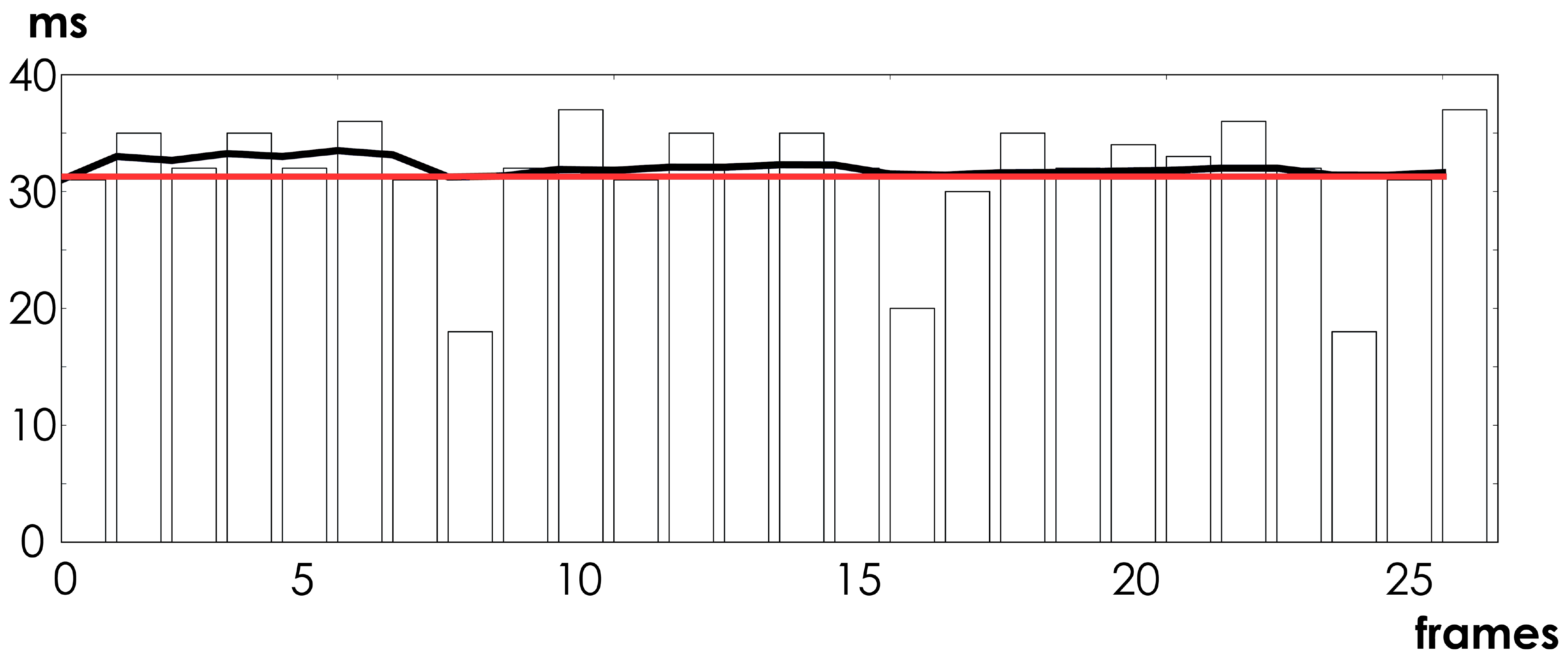}
\caption{Distances between 4-sample frames. Red line indicates expected distance of $4/127.88 = 0.03106ms$. The bars indicate the observed distance. We can see that the Emotiv system compensates every $8 \times 4 = 32$ samples to keep the average (black line) at the correct level.}
\label{figure_distance_4}
\end{figure}

In the second set of measurements, we test the stability of the timing of the packets as they appear in the system. To measure this, we collect the packets from the Emotiv EPOC device and change the screen color every 4 packets (limited by screen refresh rate, $60Hz$). This change is then measured by a photocell and fed into the oscilloscope and the distance between the 4-packet packages is calculated. Fig.~\ref{figure_distance_4} shows these measurements.
In summary, the stability and quality of the acquired signal is excellent. Most of the variations, including imperfect sampling rate or timing jitters are constant and can be largely accounted for in the data analysis if necessary.

		\subsubsection{3D Source Reconstruction On-Device Performance}
 
Source imaging was obtained using the Bayesian inverse solver for the linear model in Eq.\ (\ref{eq:ForwardProblem}). The forward matrix ${\bf A}$ and cortical source mesh grid was based on a coarse resolution (5124 vertices) of the SPM8 template brain \cite{litvak2011eeg}, further reduced to 1028 using Matlab's function reducepatch. We tested the performance of 3D reconstruction and hyper-parameters calculation on 1s of signal: MacBookPro8,2 (Intel Core i7 Sandy Bridge 2.2GHz): 2ms/2s, Nexus 7: 8ms/1s, Galaxy Note: 8ms/11s, and Acer Iconia: 14ms/13s.

	\subsection{Imagined Finger Tapping -- Online Source Reconstruction}
In order to demonstrate the applicability of discriminating a simple task as the left and right imagined finger tapping on the cortical source level in an online framework, the EEG data were acquired with the Emotiv EPOC neuroheadset and compared with EEG recordings acquired with a Biosemi Active-II device 64 channels. The 64-channels were subsampled to represent the same channel locations as the Emotiv device.

\begin{figure}
        \centering
        \begin{subfigure}[b]{0.8\columnwidth}
                \centering
        		\includegraphics[width=\textwidth]{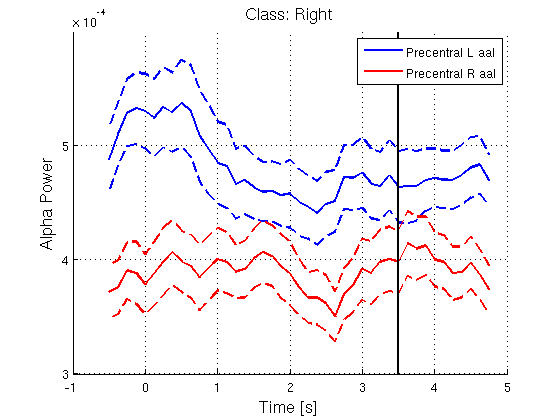}
				\caption{Emotiv EPOC.}
				\label{figure_Emotiv_PowerInPrecentralAALregions}
        \end{subfigure}%

        ~ 
        \begin{subfigure}[b]{0.8\columnwidth}
                \centering
               \includegraphics[width=\textwidth]{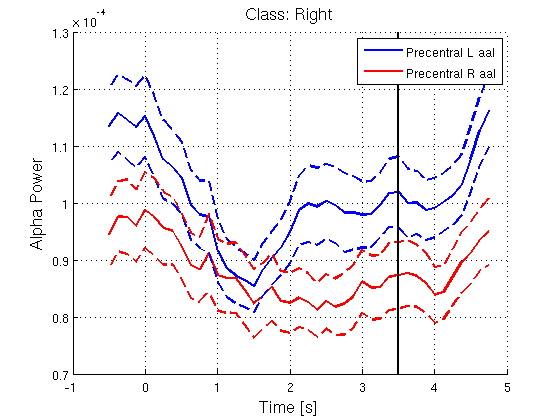}
\caption{Biosemi 14 channels. }
\label{figure_Biosemi14ch_PowerInPrecentralAALregions}
        \end{subfigure}
     \caption{Mean (solid lines) and standard deviation (dashed lines) of reconstructed current source power in the left (L) and right (R) Precentral AAL regions calculated across Right cued imagined finger tapping conditions. Online estimation of the $\alpha$ and $\beta$ parameters. Minimum Norm Solution.}
\end{figure}

Imagined fingertapping is known to lead to a suppression of the alpha (8-13 Hz) activity over the premotor/motor regions, with the contra lateral areas normally being more desynchronized \cite{pfurtscheller1999event}. Thus, imagined right finger tapping, should lead to the alpha activity being suppressed both in the left and right pre-motor region with the Left as the dominant one.
This is confirmed in Fig.~\ref{figure_Emotiv_PowerInPrecentralAALregions} and Fig.~\ref{figure_Biosemi14ch_PowerInPrecentralAALregions}, which demonstrate the SBS2 framework ability to online reconstruct meaningful current sources within the brain. Fig.~\ref{figure_Emotiv_PowerInPrecentralAALregions} shows how alpha power (8-13 Hz) is suppressed over time in the two regions of interest; Precentral Left and Right AAL. The responses are calculated as the averaged response over 87 Right cued trials. Note, that even though this result is an average over runs, the source localization was carried out in online mode with model parameters ($\alpha$ and $\beta$) and current sources ($\bf S$) estimated online. By collecting these source estimates over time we have just presented the averaged response at the end of the experiment. Similarly, Fig.~\ref{figure_Biosemi14ch_PowerInPrecentralAALregions} demonstrates the averaged power response across 79 Right cued trials.
Interestingly, the suppression of the alpha power in the Left and Right Precentral AAL regions to right imagined finger tapping trials, looks quite similar for both devices (Emotiv EPOC and Biosemi), with the contralateral frontal regions (Left) being mostly suppressed.

\section{Conclusions}

We have presented the design, implementation, and evaluation of the first fully mobile 3D EEG imaging system: The \emph{Smartphone Brain Scanner}. The open source software allows realtime EEG data acquisition and source imaging on standard off-the-shelf Android mobile smartphones and tablets with a good spatial resolution and frame rates in excess of 40 fps. In particular, we have implemented a real-time solver for the ill-posed inverse problem with online Bayesian optimization of hyper-parameters (noise level and regularization).

The evaluation showed that the combined system provides for a stable imaging pipeline with a delay of 80-120ms. We showed results of a cue imagined finger tapping experiment and compared the smartphone brain scanner's average power in the alpha band in a relevant motor area, and we found that these aggregate signals compare favorably with those obtained with standard laboratory equipment. Both show the expected de-synchronization on initiation of imagined motor actions.

We suggest that the mobility and simplified application development may enable completely new research directions for imaging neuroscience and thus offset the expected reduced signal quality of a mobile off-the-shelf, low-density neuroheadset relative to more conventional and controlled, high-density laboratory equipment.

\section*{Acknowledgment}
This work is supported in part by the Danish Lundbeck Foundation through CIMBI, Center for Integrated Molecular Brain Imaging.

\bibliography{bibliography}

\end{document}